\documentclass[preprint2]{aastex}
\usepackage{graphicx}

\newcommand{\teff}{\mbox{$T_{\rm eff}$}}
\newcommand{\logg}{\mbox{$\log g$}}
\newcommand{\vsini}{\mbox{$v \sin i$}}
\newcommand{\mictrb}{\mbox{$\xi_{\rm t}$}}
\newcommand{\mactrb}{\mbox{$v_{\rm mac}$}}

\newcommand{\kms}{\mbox{km\,s$^{-1}$}}
\newcommand{\halpha}{\mbox{$H_\alpha$}}

\slugcomment{Based on data collected with the HARPS spectrograph at
ESO La Silla Observatory for program 082.C-0608(A).}

\shorttitle{WASP-22\,b}
\shortauthors{Maxted et~al.}

\begin{document} 
\title {WASP-22\,b: A transiting ``hot Jupiter'' planet in a hierarchical triple
system}
\author{P.~F.~L.~Maxted\altaffilmark{1},
D.~R.~Anderson\altaffilmark{1},
M.~Gillon\altaffilmark{2,3},
C.~Hellier\altaffilmark{1},
D.~Queloz\altaffilmark{2},
B.~Smalley\altaffilmark{1},
A.~H.~M.~J.~Triaud\altaffilmark{2},
R.~G.~West\altaffilmark{4},
D.~M.~Wilson\altaffilmark{1},
S.~J.~Bentley\altaffilmark{1},
H.~Cegla\altaffilmark{1,5,9}
A.~Collier~Cameron\altaffilmark{6},
B.~Enoch\altaffilmark{6},
L.~Hebb\altaffilmark{6},
K.~Horne\altaffilmark{6},
J.~Irwin\altaffilmark{7},
T.~A.~Lister\altaffilmark{8},
M.~Mayor\altaffilmark{2},
N.~Parley\altaffilmark{6},
F.~Pepe\altaffilmark{2},
D.~Pollacco\altaffilmark{9},
D.~Segransan\altaffilmark{2},
S.~Udry\altaffilmark{2},
P.~J.~Wheatley\altaffilmark{10},
}

\altaffiltext{1}{Astrophysics Group, Keele University, Staffordshire, ST5 5BG, UK}
\altaffiltext{2}{Observatoire de Gen\`eve, Universit\'e de Gen\`eve, 51 Chemin des Maillettes, 1290 Sauverny, Switzerland}
\altaffiltext{3}{Institut d'Astrophysique et de G\'eophysique,  Universit\'e de Li\`ege,  All\'ee du 6 Ao\^ut, 17,  Bat.  B5C, Li\`ege 1, Belgium}
\altaffiltext{4}{Department of Physics and Astronomy, University of Leicester, Leicester, LE1 7RH, UK}
\altaffiltext{5}{Physics \& Astronomy Department, Vanderbilt University, 6301
Stevenson Center, Nashville, TN 37235, USA}
\altaffiltext{6}{School of Physics and Astronomy, University of St. Andrews, North Haugh, Fife, KY16 9SS, UK}
\altaffiltext{7}{Department of Astronomy, Harvard University, 60 Garden Street, MS 10, Cambridge, Massachusetts 02138, USA}
\altaffiltext{8}{Las Cumbres Observatory, 6740 Cortona Dr. Suite 102, Santa Barbara, CA 93117, USA}
\altaffiltext{9}{Astrophysics Research Centre, School of Mathematics \& Physics, Queen's University, University Road, Belfast, BT7 1NN, UK}
\altaffiltext{10}{Department of Physics, University of Warwick, Coventry, CV4 7AL, UK}

\begin{abstract}
 We report the discovery of a transiting planet orbiting the star
TYC~6446-326-1. The star, WASP-22, is a moderately bright (V=12.0) solar-type
star ($\teff=6000\pm 100$\,K, [Fe/H]$ = -0.05\pm 0.08$). The lightcurve of the
star obtained with the WASP-South instrument shows periodic transit-like
features with a depth of about 1\% and a duration of 0.14\,d.
 The presence of a transit-like feature in the lightcurve is
confirmed using z-band photometry obtained with  Faulkes Telescope South. High
resolution spectroscopy obtained with the CORALIE  and HARPS spectrographs
confirm the presence of a planetary mass companion with an orbital period of
3.533\,days in a near-circular orbit. From a combined analysis of
the spectroscopic and photometric data assuming that the star is a typical
main-sequence star we estimate that the planet has a mass $M_{\rm p} =
0.56\pm 0.02M_{\rm Jup}$ and a radius $R_{\rm p} = 1.12 \pm 0.04R_{\rm Jup}$.
In addition, there is a linear trend of 40\,m\,s$^{-1}$\,y$^{-1}$ in the
radial velocities measured over 16 months, from which we infer the presence of
a third body with a long period orbit in this system. The companion may be a
low mass M-dwarf, a white dwarf or a second planet. 

\end{abstract}


\section{Introduction}
 
  The WASP project \citep{2006PASP..118.1407P} is currently one of the most
successful wide-area surveys designed to find exoplanets transiting 
bright stars (V $<$ 12.5).  Other succesful surveys include HATnet
\citep{2004PASP..116..266B}, XO \citep{2005PASP..117..783M} and TrES
\citep{2006AAS...20922602O}. There is continued interest in finding transiting
exoplanets because they can be accurately characterized and studied in some
detail, e.g., the mass and radius of the planet can be accurately measured.
This gives us the opportunity to explore the relationships between the density
of the planet and other properties of the planetary system, e.g., the
semi-major axis, the spectral type of the star, the eccentricity of the orbit,
etc. Given the wide variety of transiting planets being discovered and the
large number of parameters that characterize them, statistical studies will
require a large sample of systems to identify and quantify the relationships
between these parameters. These relationships can be used to test models of
the formation, structure and evolution of short period exoplanets.

 A particular puzzle related to the properties of hot Jupiters is the wide
range in their densities. Very dense hot Jupiters such as HD~149026 are
thought to contain a dense, metallic core \citep{2005ApJ...633..465S}. There
is currently no generally agreed explanation for the existence of hot Jupiters
with densities 5\,--\,10 times lower than the density of Jupiter, e.g.
WASP-17\,b \citep{2010ApJ...709..159A}, TrES-4\,b \citep{2007ApJ...667L.195M}
and WASP-12\,b \citep{2009ApJ...693.1920H}. One possibility is that the
planets are heated by tidal forces, and that these are driven by the presence
of a third body in the system \citep{2007MNRAS.382.1768M}. Other possibilities
include  enhanced opacity in the atmosphere \citep{2007ApJ...661..502B}, the
distribution of heavy elements in the core \citep{2008A&A...482..315B} and
kinetic heating from the irradiated atmosphere into the interior
\citep{2002A&A...385..166S}.

 Here we report the discovery of a hot Jupiter system, WASP-22, identified
using the WASP-South instrument and present evidence that it is a member of a
hierarchical triple system.

\begin{table} 
\caption{Radial velocity measurements.} 
\label{rv-data} 
\begin{tabular*}{0.5\textwidth}{@{\extracolsep{\fill}}lrrr} 
\tableline 
\noalign{\smallskip}
BJD--2\,400\,000 &\multicolumn{1}{l}{RV} & 
\multicolumn{1}{l}{$\sigma_{\rm RV}$} & 
\multicolumn{1}{l}{BS}\\ 
 & (km s$^{-1}$) & (km s$^{-1}$) & (km s$^{-1}$)\\ 
\noalign{\smallskip}
\tableline
\noalign{\smallskip}
CORALIE&&&\\
54704.8563 &$ -7.236 $& 0.013 & $0.005$\\
54706.8761 &$ -7.368 $& 0.026 & $0.030$\\
54708.8417 &$ -7.189 $& 0.011 & $-0.020$\\
54709.7625 &$ -7.320 $& 0.019 & $-0.038$\\
54710.7697 &$ -7.291 $& 0.024 & $0.046$\\
54715.8843 &$ -7.201 $& 0.018 & $0.058$\\
54716.8457 &$ -7.262 $& 0.041 & $-0.127$\\
54717.7794 &$ -7.386 $& 0.041 & $0.034$\\
54720.7520 &$ -7.312 $& 0.017 & $0.002$\\
54721.8184 &$ -7.289 $& 0.019 & $0.028$\\
54722.8233 &$ -7.201 $& 0.012 & $-0.033$\\
54724.7367 &$ -7.314 $& 0.017 & $-0.011$\\
54726.8108 &$ -7.225 $& 0.012 & $0.010$\\
54729.8623 &$ -7.206 $& 0.013 & $-0.005$\\
54731.8658 &$ -7.345 $& 0.012 & $-0.015$\\
54740.7564 &$ -7.199 $& 0.020 & $-0.058$\\
54834.5647 &$ -7.304 $& 0.018 & $0.014$\\
54836.5707 &$ -7.225 $& 0.017 & $-0.019$\\
54853.6313 &$ -7.171 $& 0.014 & $0.046$\\
54854.6322 &$ -7.275 $& 0.013 & $0.018$\\
54855.6091 &$ -7.310 $& 0.012 & $0.001$\\
54860.5404 &$ -7.183 $& 0.013 & $0.051$\\
54862.5589 &$ -7.327 $& 0.015 & $-0.046$\\
54865.5628 &$ -7.313 $& 0.014 & $0.024$\\
54879.5256 &$ -7.280 $& 0.017 & $-0.039$\\
54880.5402 &$ -7.273 $& 0.021 & $0.033$\\
54882.5242 &$ -7.193 $& 0.021 & $0.043$\\
54885.5207 &$ -7.159 $& 0.016 & $-0.014$\\
54886.5443 &$ -7.299 $& 0.017 & $-0.015$\\
55095.8758 &$ -7.308 $& 0.013 & $0.013$\\
55100.8593 &$ -7.154 $& 0.015 & $-0.002$\\
55126.7935 &$ -7.266 $& 0.015 & $0.005$\\
55128.7415 &$ -7.169 $& 0.016 & $0.006$\\
55185.6618 &$ -7.147 $& 0.011 & $-0.007$\\
55186.6153 &$ -7.234 $& 0.011 & $-0.025$\\
55189.6571 &$ -7.164 $& 0.010 & $0.004$\\
55190.6690 &$ -7.284 $& 0.011 & $-0.008$\\
HARPS&&&\\                     
54743.7527 & $-7.1832 $& 0.0029 &$ 0.0090$\\
54746.7683 & $-7.2391 $& 0.0026 &$ 0.0040$\\
54749.7650 & $-7.2941 $& 0.0024 &$ 0.0039$\\
54750.6791 & $-7.1940 $& 0.0029 &$ 0.0109$\\
54754.7257 & $-7.1752 $& 0.0025 &$ 0.0118$\\
54755.7558 & $-7.2726 $& 0.0032 &$ 0.0329$\\
\noalign{\smallskip}
\tableline 
\end{tabular*} 
\end{table}

\begin{figure} 
\plotone{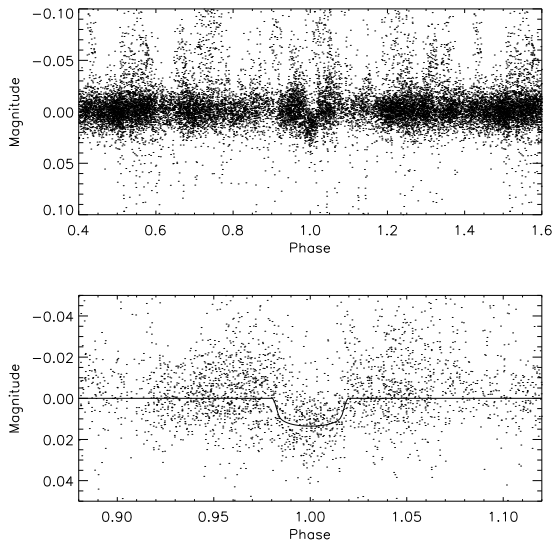}
\caption{WASP-South photometry of WASP-22 folded on the orbital period
$P$~=~3.532759\,d. Upper panel: all data. Lower panel: data within 0.12 phase
units of mid-transit together with the model fit described in
Section~\ref{paramsec} (solid line). \label{phot-full-and-zoom} }
\end{figure} 
\begin{figure} 
\plotone{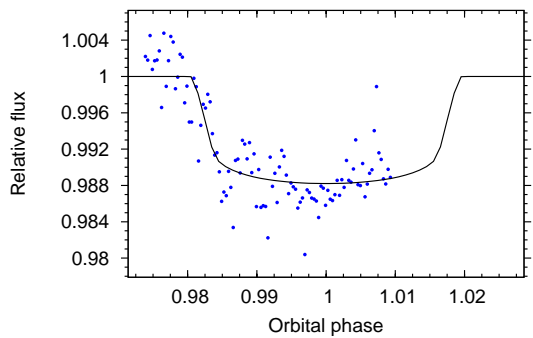} 
\caption{Faulkes Telescope South z-band photometry of WASP-22 (points) with 
 the model fit described in
Section~\ref{paramsec} (solid line). 
\label{ftslc} }
\end{figure}

\section{Observations}
The WASP survey is described in \citet{2006PASP..118.1407P}
and \citet{2008ApJ...675L.113W} while a discussion of our candidate
selection methods can be found in \citet{2007MNRAS.380.1230C},
\citet{2008MNRAS.385.1576P}, and references therein. 

 The WASP-South instrument consists of 8 cameras, each with a Canon 200-mm
f/1.8 lens and a 2k$\times$2k $e$2$V$ CCD detector  resulting in an image
scale of approximately 14\,arcsec per pixel.  The star TYC~6446-326-1 (=
1SWASP J033116.32$-$234911.0) was observed 3133 times by one camera on the
WASP-South instrument from 2006 August to 2007 January. A further 6282
obervations were obtained with the same camera from 2007 August to 2008
January. The star also appeared in the images obtained with a second camera
during the second observing season, so a further 5889 observations were
obtained with this camera during that interval.

\begin{table}
\begin{center}
\caption{Stellar parameters of WASP-22 from our spectroscopic analysis.}
\begin{tabular}{lr}
\tableline\tableline
Parameter & Value \\
\noalign{\smallskip}
\tableline
\teff~(K)  & 6000 $\pm$ 100 \\
\logg      & 4.5 $\pm$ 0.2 \\
\mictrb~(\kms)    & 1.2 $\pm$ 0.1 \\
\vsini ~ (\kms)    & 3.5 $\pm$ 0.6 \\
{[Fe/H]}   &$-$0.05 $\pm$ 0.08 \\
{[Mg/H]}   &  +0.06 $\pm$ 0.04 \\
{[Si/H]}   &  +0.17 $\pm$ 0.11 \\
{[Ca/H]}   &  +0.10 $\pm$ 0.11 \\
{[Sc/H]}   &  +0.12 $\pm$ 0.12 \\
{[Ti/H]}   &  +0.10 $\pm$ 0.07 \\
{[V/H]}    &  +0.07 $\pm$ 0.09 \\
{[Cr/H]}   &\phantom{+}0.00 $\pm$ 0.08 \\
{[Mn/H]}   &  +0.02 $\pm$ 0.06 \\
{[Co/H]}   &  +0.09 $\pm$ 0.13 \\
{[Ni/H]}   &  +0.09 $\pm$ 0.06 \\
log N(Li)  &   2.23 $\pm$ 0.08 \\
\tableline
\end{tabular}
\label{wasp22-params}
\end{center}
\end{table}

 The WASP-South lightcurves of WASP-22 show transit-like features with a depth
of approximately 0.012 magnitudes recurring with a 3.53-d period
(Fig.~\ref{phot-full-and-zoom}).  These were detected in the WASP
photometry from the 2006 season using the de-trending and transit detection
methods described in \cite{2007MNRAS.380.1230C}, but judged to be
``unconvincing'' at that stage. The same 3.53-d periodicity was detected in
the 2007 data  from both cameras, which was taken as  strong evidence that the
periodic transit signal was real. The spectral type of the star was estimated
to be approximately G1 based on the catalogue photometry available for this
star at the time. The duration and depth of the transit is consistent with the
hypothesis that it is due to the transit of a planet-like companion
to a main-sequence G1 star and the WASP lightcurves show no indication of any
ellipsoidal variation due to the distortion of the star by a massive
companion.

 We also considered whether WASP-22 was likely to be a dwarf star given its
J$-$H colour and J-band reduced proper motion, H$_{\rm J}$. The position of
WASP-22 relative to the polynomial boundary between dwarfs and giants given in
\cite{2007MNRAS.380.1230C} suggests that WASP-22 is a dwarf.  We have also
used the results of the RAVE survey DR2 \citep{2008AJ....136..421Z} to calibrate
the probability that a star is a dwarf given the observed values of J$-$H and
H$_{\rm J}$. The RAVE survey observed stars at a similar range of magnitudes
and galactic positions to the WASP survey and so it is well suited to this
purpose. Stars listed in the survey with $\logg \ge 3.5$ were identified as
dwarf stars, those with lower $\logg$ estimates were identified as giants. We
then created a look-up table of the relative numbers of dwarfs and giants as a
function of position in the $H_{\rm J}$ v. J$-$H plane.  This criterion also
suggested that WASP-22 is likely to be a dwarf star because the ratio of
dwarfs to giants in this region of the $H_{\rm J}$ v. J$-$H plane is
dwarf:giant = 10:0.

  The star was included as a ``grade-A'' candidate  in our programme of
follow-up spectroscopic observations using the CORALIE spectrograph on the
Euler 1.2-m telescope and the HARPS spectrograph on the 3.6-m ESO telescope,
both located at La Silla, Chile. We obtained 37 radial-velocity measurements
during the interval 2008 August 27 to 2009 December 25 with CORALIE and 6
measurements with HARPS during the interval 2008 Oct 4\,--\,16.  The
CORALIE spectra have a signal-to-noise ratio of 10\,--\,20, while the typical
signal-to-noise ratio of the HARPS spectra is 50. The measurements are given
in Table~\ref{rv-data}, where we also provide the bisector span, BS, which
measures the asymmetry of the cross-correlation function. The standard error
of the the bisector span measurements is $2\sigma_{\rm RV}$. 

 We also obtained photometry of  TYC~6446-326-1 and other nearby stars on 2009
August 8 using the LCOGT 2.0-m Faulkes Telescope South (FTS) at Siding Spring
Observatory. The Merope camera we used has an image scale of 0.279
arcseconds/pixel when used in the 2x2 binning mode we employed. We used a
Pan-STARRS\footnote{\url{http://pan-starrs.ifa.hawaii.edu/public}}
z-band filter to obtain 107 images covering one ingress of the transit. These
images were processed in the standard way with IRAF\footnote{IRAF is
distributed by the National Optical Astronomy Observatory, which is operated
by the Association of Universities for Research in Astronomy (AURA) under
cooperative agreement with the National Science Foundation.} using a stacked
bias image, dark frame, and sky flat. Minimal fringing was present in the
z-band images due to the deep depletion CCD in the camera, so no fringe
correction was applied. The DAOPHOT photometry package
\citep{1987PASP...99..191S} was used to perform object detection and aperture
photometry with an aperture size of 10 binned pixels in radius. The
5'\,$\times$\,5' field of view of the instrument contained 6 comparison stars
that were used in deriving the differential magnitudes with a photometric
precision of 3.1 mmag. The coverage of the out-of-transit phases is quite
limited, but the data are sufficient to confirm that transit-like features
seen in the WASP-South data are due to the star TYC~6446-326-1 and to provide
better measurements of the depth of the transit and the duration of ingress
than is possible from the WASP-South data (Fig.~\ref{ftslc}). We note that the
star 1SWASP~J033121.40-234857.8 77\,arcsec from WASP-22 is also a variable
star. This star showed a dip in brightness of about 0.01\,magnitudes lasting
about 100\,minutes during our observation of WASP-22.  

  We examined the power spectra of the WASP data from each camera and season
separately to look for any intrinsic photometric variability due to magnetic
activity in WASP-22. There is no periodic signal consistently detected in all
three data sets so any such variability must have an amplitude less than a few
milli-magnitudes.
 
 All photometric presented in this paper are available from the NStED
database.\footnote{\url{http://nsted.ipac.caltech.edu}}

\begin{figure} 
\plotone{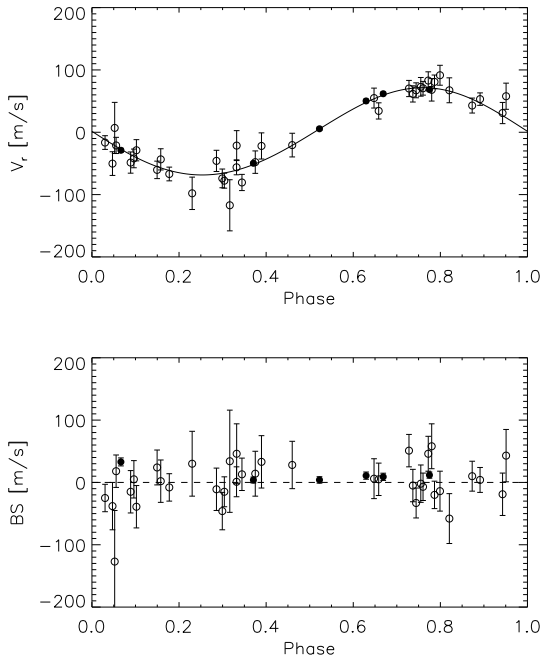} 
\caption{Radial velocity and bisector span measurements for WASP-22. Upper
panel: Radial velocity relative to the centre-of-mass velocity including the
long-term drift $\frac{d\gamma}{dt}$ (points with error bars) compared to our
model for the spectroscopic orbit (solid line).  Lower panel: bisector span
measurements.  Data obtained with the HARPS spectrograph
are plotted with filled circles, CORALIE data with open circles.
\label{rvphase} }
\end{figure} 

\begin{figure} 
\plotone{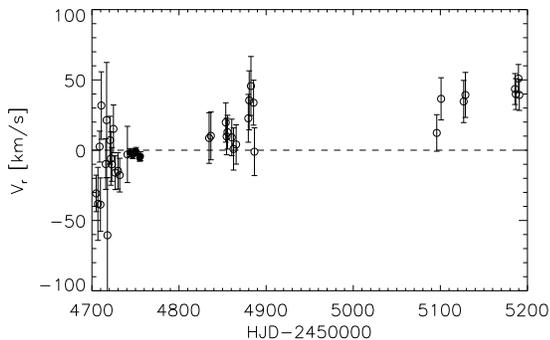} 
\caption{Residuals from a Keplerian orbit fit to our radial velocity data
without a linear trend in the model. Symbols are as for Fig.~\ref{rvphase}.
\label{rvtime} }
\end{figure}

\section{WASP-22 Stellar Parameters }

 The individual HARPS spectra of WASP-22 were co-added
to produce a single spectrum with a typical S/N of around 100:1 that we have
analysed to determine the atmospheric parameters of the star. The standard
pipeline reduction products were used in the analysis.

The analysis was performed using the {\sc uclsyn} spectral synthesis package
\citep{1992PhDT.........1S,uclsyn} and {\sc atlas9} models without convective
overshooting \citep{1997A&A...318..841C}. The \halpha\ line was used to
determine the effective temperature (\teff), while the Na {\sc i} D and Mg
{\sc i} b lines were used as surface gravity (\logg) diagnostics. The
parameters obtained from the analysis are listed in Table~\ref{wasp22-params}.

The equivalent widths of several clean and unblended lines were measured.
Atomic line data was mainly taken from the \citet{1995all..book.....K}
compilation, but with updated van der Waals broadening coefficients for lines
in \citet{2000A&AS..142..467B} and $\log gf$ values from
\citet{2000AJ....119..390G},
\citet{2001AJ....121..432G} or \citet{2004A&A...415.1153S}. A value for microturbulence
(\mictrb) was determined from Fe~{\sc i} using the method of
\citet{1984A&A...134..189M}. The
ionization balance between Fe~{\sc i} and Fe~{\sc ii} and the null-dependence
of abundance on excitation potential were used as an additional \teff\ and
\logg\ diagnostics \citep{2005MSAIS...8..130S}.

 We have determined the elemental abundances of several elements (also listed in
Table~\ref{wasp22-params}) from their measured equivalent widths. The quoted
error estimates include that given by the uncertainties in \teff, \logg\ and
\mictrb, as well as the scatter due to measurement and atomic data
uncertainties. There is no evidence from these data for any abundance
anomalies in WASP-22.

 The spectrum of WASP-22 shows a clear Li {\sc i} 6708\AA\ line
with EW = 32 $\pm$ 1 m\AA, indicating an abundance of {log n(Li/H) + 12} $=$
2.23 $\pm$ 0.08 dex. We have compared this lithium abundance to the relations
between effective temperature and age by 
\citet{2005A&A...442..615S}. For stars with $\teff\approx 6000$\,K we find that
this lithium abundance implies a lower limit to the age of about 1\,Gyr.

The projected stellar rotation velocity (\vsini) was determined by fitting the
profiles of several unblended Fe~{\sc i} lines. A value for macroturbulence
(\mactrb) of 4.5 \kms\ was assumed, based on the tabulation by
\citet{2008oasp.book.....G}, and an instrumental FWHM of 0.065\AA, determined
from the telluric lines around 6300\AA. A best fitting value of \vsini\ = 3.5
$\pm$ 0.6~\kms\ was obtained.  Inspection of the Ca\,II H and K lines
shows no indication of any emission due to chromospheric activity, as expected
for such a slowly rotating star. If we assume that the rotation axis of the
star is approximately aligned with the orbital axis of the planet's orbit we
can estimate that the rotation period of the star is $P_{\rm rot} = 16\pm
3$\,d. The rotation-colour-age relation of \cite{2009MNRAS.400..451C} then
yields an age estimate of approximately $3\pm 1$\,Gyr for this star. We have
also compared the values of \teff\ and the stellar density $\rho_{\star}$ to
the models of \cite{2000A&AS..141..371G} and find that the age of the star is
likely to be less than 6\,Gyr based on these models.

 We did not see any indication of additional spectral lines in the spectrum
nor any trend in equivalent widths that might suggest contamination of the
spectrum by another star. We estimate that the third body in this system
discussed below contributes less than 5\% of the light in the optical
spectrum. 

 In summary, WASP-22 appears to be a slowly rotating, chromospherically
inactive, main-sequence star that is slightly hotter than the Sun. 

\subsection{Planetary parameters\label{paramsec}}

 The amplitude of the radial velocity variation with the same period as the
transit lightcurve (Fig.~\ref{rvphase}) and the lack of any correlation
between this variation and the bisector span establish the presence of a
planetary mass companion to this star \citep{2001A&A...379..279Q}.

 The CORALIE and HARPS radial velocity measurements were combined with the
WASP-South and FTS photometry in a simultaneous Markov-chain Monte-Carlo
(MCMC) analysis to find the parameters of the WASP-22 system. The shape of the
transit is not well defined in the WASP-South or FTS photometry, so we have
imposed an assumed main-sequence mass-radius relation as an additional
constraint in our analysis of the data. The  stellar mass is determined from
the parameters \teff, \logg\ and [Fe/H] using the procedure described by
\cite{2010arXiv1004.1991E}, based on the compilation of eclipsing binary data by
\cite{2010A&ARv..18...67T}. The code uses \teff\ and [Fe/H] as MCMC jump
variables, constrained by Bayesian priors based on the
spectroscopically-determined values given in Table~\ref{wasp22-params}. 
Limb-darkening coefficients are taken from \cite{2000A&A...363.1081C}.

  A careful analysis of the radial velocity data clearly shows a trend in the
residuals as a function of time (Fig.~\ref{rvtime}). We therefore adapted our
MCMC analysis to include an extra parameter, $\frac{d\gamma}{dt}$, which
describes a linear trend in the centre-of-mass velocity of the star. 

 The parameters derived from our MCMC analysis are listed in
Table~\ref{sys-params}. We found that we did not need to account for any
additional noise in the radial velocity data to due to stellar activity
(``jitter''). The contribution to the total chi-squared from our 43 radial
velocity measurements is 32.4. The surface gravity derived from our MCMC
solution is consistent with the \logg\ value from the analysis of the
spectrum, but the large uncertainty on the latter value means that this is a
rather weak constraint.

 We considered the contribution of
red noise to the standard errors quoted in Table~\ref{sys-params}.  While the
individual transits in the WASP data are affected by red noise, the analysis
of the combined lightcurve covering many individual transits will not be
strongly affected by red noise because there will be no correlation between
the systematic noise from different nights. The FTS lightcurve is affected by
red noise so we have investigated the effect of this using the ``prayer bead''
method. A separate MCMC analysis was performed in which synthetic FTS
lightcurves were created from the model fit to the lightcurve and the residuals
from this model after cyclic permutation at each step in the MCMC chain. We
find that this does not significantly increase the error estimates for any of
the parameters with the exception of the transit epoch and period. The
standard errors for these two parameters in Table~\ref{sys-params} are taken
from the ``prayer bead'' MCMC analysis.

 We have included the parameters $e\sin(\omega)$ and $e\cos(\omega)$ as free
parameters in the solution because the orbit of the planet is not known to be
circular, although the values of derived and their standard errors are
quite consistent with this hypothesis. Imposing a circular orbit in the
solution has a negligible effect on the value of the other parameters derived,
but would reduce the estimated errors on these parameters, perhaps
unrealisitically so.

\section{Discussion}
Apart from the presence of a third body, WASP-22 is a typical hot Jupiter
planetary system. WASP-22\,b joins a growing number of planets discovered with
masses $\approx 0.5M_{\rm Jup}$, radii $\approx 1R_{\rm Jup}$ orbiting
solar-like stars with periods of about 3 days, e.g., WASP-13\,b,
WASP-11\,b/HAT-P-10\,b, CoRoT-5\,b, WASP-6\,b, HAT-P-1\,b, OGLE-TR-111\,b,
WASP-25\,b, HAT-P-3\,b, Kepler-8\,b, etc.\footnote{\it
http://exoplanet.eu}
 
 The third body we have detected from the linear trend in the radial
velocities contributes less than 5\% of the flux in the optical spectrum, so
it is at least 3 magnitudes fainter than WASP-22 at V. This rules out the
possibilty that the companion is K-type dwarf star, but leaves the possibility
that the companion is an  M-dwarf or a white dwarf.  There is good agreement
between the effective temperature we derive from the analysis of the optical
spectrum and the effective temperature derived using the infrared flux method
\citep{1979MNRAS.188..847B}. We estimate that we would have been able detect a
cool companion to WASP-22 if the 2MASS K$_s$-band magnitude (K$_s$ =
10.318$\pm$0.020, \citealt{2006AJ....131.1163S}) were increased by more than
0.1 magnitudes. This also rules out a K-dwarf companion but not an M-dwarf
companion or a white dwarf companion.  For example, a K-dwarf companion to
WASP-22 would produce K-band infrared excess of 0.1 magnitudes or more, in
which case the effective temperature derived using the infrared flux method
would be inconsistent with the value derived from the analysis of the optical
spectrum.

 There are few direct constraints on the properties of the third body from the
radial velocity data available to date. The observation that the trend is
linear over 16 months suggests that the orbital period is at least a few years.
If the orbit of the third body is approximately circular and coplanar with the
inner orbit with a period of several years and an amplitude comparable to the
velocity range observed so far, then it is possible that the third body is a
second planet, i.e., a configuration  similar to the double-planet system
HAT-P-13 \citep{2009ApJ...707..446B}. Further radial velocity monitoring will
be required to determine whether any or all of these assumptions is
reasonable. This is certainly worth doing because the detection of a second
planet in the  WASP-22 system may make it possible to infer the interior
structure of WASP-22\,b using the method of \cite{2009ApJ...704L..49B}. This
method relies on an accurate measurement of the  eccentricity for the orbit of
the inner planet. The data available to date are only of sufficient quality to
state that the eccentricity of the orbit is small ($e \la 0.06$) so more data
of the quality we obtain from HARPS will be required to measure a precise
value for the eccentricity of WASP-22\,b's orbit.

 The distance to WASP-22 is approximately 300pc so it may be possible to
detect the companion directly if it is a M-dwarf using high-constrast,
high-resolution imaging.
 
\section{Conclusion}
 The star WASP-22 (TYC~6446-326-1) has a hot Jupiter companion.  A long-term
linear trend in the mean value of the radial velocity shows that WASP-22 has a
distant companion, i.e., it is a hierarchical triple system. The properties of
the third body are poorly constrained by the data available to-date, but it
may be an M-dwarf, a white dwarf or a second planet.

\acknowledgments
WASP-South is hosted by the South African Astronomical Observatory and we are
grateful for their ongoing support and assistance. Funding for WASP comes from
consortium universities and from the UK’s Science and Technology Facilities
Council.

\begin{table*} 
\begin{center} 
\caption{System parameters for WASP-22. The
planet equilibrium temperature is calculated assuming a value for the Bond
albedo A$=0$. {\bf N.B.} an assumed main-sequence mass-radius relation is
imposed as an additional constraint in this solution so the mass and radius of
the star are not independent parameters -- see \cite{2010arXiv1004.1991E}) for
details.  } 
\label{sys-params} 
\begin{tabular}{llrl} 
\tableline\tableline
Parameter & \multicolumn{1}{l}{Symbol} & \multicolumn{1}{l}{Value} & Unit \\ 
\tableline 
Transit epoch (BJD) & $T_{\rm C}$ & $ 2454780.2496 \pm 0.0042$ & d\\
Orbital period & $P$ &$ 3.53269 \pm 0.00004$ & d\\
Transit duration & $T_{\rm 14}$ & $ 0.137 \pm 0.003  $ & d\\
Planet/star area ratio & $R_{\rm P}^{2}$/R$_{*}^{2}$ & $ 0.0104 \pm 0.0004 $ & ~\\
Impact parameter & $b$ & 0.13$\pm 0.08$  &  \\
Stellar reflex velocity & K$_{\rm 1}$ & $ 70.0 \pm 1.7   $ & m\,s$^{-1}$\\
Centre-of-mass velocity at time $T_{\rm C}$ & $\gamma_{\rm CORALIE}(T_{\rm C})$ &$ -7262 \pm
2 $ & m\,s$^{-1}$\\
Velocity offset, HARPS\,$-$\,CORALIE & $\gamma_{\rm HARPS} - \gamma_{\rm
CORALIE}$ & $ 21 \pm 2 $ & m\,s$^{-1}$\\

Drift in centre-of-mass velocity & $\frac{d\gamma}{dt}$ & $40 \pm 5$ &
m\,s$^{-1}$\,yr$^{-1}$\\
Orbital separation & $a$ & $ 0.0468 \pm 0.0004 $ & AU \\
Orbital inclination & $i$ & 89.2 $\pm 0.5$ & $^{\circ}$ \\
Orbital eccentricity & $e$ & $0.023 \pm 0.012$ & ~\\
Arg. of periastron & $\omega$ &  27$^{+ 51}_{- 78}$ & $^{\circ}$ \\
 ~ & $e\cos(\omega)$ & $0.012 \pm 0.011 $\\
 ~ & $e\sin(\omega)$ & $0.006 \pm 0.021 $\\
\noalign{\smallskip}
\noalign{\smallskip}
Stellar mass & $M_{\rm *}$ & $ 1.1 \pm 0.3 $ & M$_{\rm \sun}$\\
Stellar radius & $R_{\rm *}$ &$ 1.13 \pm 0.03$ & R$_{\rm \sun}$\\
Stellar surface gravity & $\log g_{*}$ & $ 4.37 \pm 0.02 $  & (cgs)\\
Stellar density & $\rho_{\rm *}$ & 0.76$\pm 0.06$ & $\rho_{\rm \sun}$ \\
Planet mass & $M_{\rm P}$ & $ 0.56 \pm 0.02 $ & M$_{\rm J}$\\
Planet radius & $R_{\rm P}$ & $1.12 \pm 0.04 $ & R$_{\rm J}$\\
Planet surface gravity & $\log g_{\rm P}$ & $3.00 \pm 0.03 $ & (cgs)\\
Planet density & $\rho_{\rm P}$ & $ 0.40 \pm 0.04 $ & $\rho_{\rm J}$\\
Planet equil. temp.  & $T_{\rm P}$ &$1430 \pm 30 $ & K\\
\tableline 
\end{tabular} 
\end{center} 
\end{table*}

\bibliographystyle{pasp}
\bibliography{wasp}

\begin{thebibliography}{}

\bibitem[\protect\citeauthoryear{{Anderson} et~al.}{{Anderson}
  et~al.}{2010}]{2010ApJ...709..159A}
{Anderson}, D.~R. et~al. 2010, \apj, 709, 159

\bibitem[\protect\citeauthoryear{{Bakos} et~al.}{{Bakos}
  et~al.}{2004}]{2004PASP..116..266B}
{Bakos}, G., {Noyes}, R.~W., {Kov{\'a}cs}, G., {Stanek}, K.~Z., {Sasselov},
  D.~D.,  and {Domsa}, I. 2004, \pasp, 116, 266

\bibitem[\protect\citeauthoryear{{Bakos} et~al.}{{Bakos}
  et~al.}{2009}]{2009ApJ...707..446B}
{Bakos}, G.~{\'A}. et~al. 2009, \apj, 707, 446

\bibitem[\protect\citeauthoryear{{Baraffe}, {Chabrier}, and {Barman}}{{Baraffe}
  et~al.}{2008}]{2008A&A...482..315B}
{Baraffe}, I., {Chabrier}, G.,  and {Barman}, T. 2008, \aap, 482, 315

\bibitem[\protect\citeauthoryear{{Barklem}, {Piskunov}, and {O'Mara}}{{Barklem}
  et~al.}{2000}]{2000A&AS..142..467B}
{Barklem}, P.~S., {Piskunov}, N.,  and {O'Mara}, B.~J. 2000, \aaps, 142, 467

\bibitem[\protect\citeauthoryear{{Batygin}, {Bodenheimer}, and
  {Laughlin}}{{Batygin} et~al.}{2009}]{2009ApJ...704L..49B}
{Batygin}, K., {Bodenheimer}, P.,  and {Laughlin}, G. 2009, \apjl, 704, L49

\bibitem[\protect\citeauthoryear{{Blackwell}, {Shallis}, and
  {Selby}}{{Blackwell} et~al.}{1979}]{1979MNRAS.188..847B}
{Blackwell}, D.~E., {Shallis}, M.~J.,  and {Selby}, M.~J. 1979, \mnras, 188,
  847

\bibitem[\protect\citeauthoryear{{Burrows} et~al.}{{Burrows}
  et~al.}{2007}]{2007ApJ...661..502B}
{Burrows}, A., {Hubeny}, I., {Budaj}, J.,  and {Hubbard}, W.~B. 2007, \apj,
  661, 502

\bibitem[\protect\citeauthoryear{{Castelli}, {Gratton}, and
  {Kurucz}}{{Castelli} et~al.}{1997}]{1997A&A...318..841C}
{Castelli}, F., {Gratton}, R.~G.,  and {Kurucz}, R.~L. 1997, \aap, 318, 841

\bibitem[\protect\citeauthoryear{{Claret}}{{Claret}}{2000}]{2000A&A...363.1081%
C}
{Claret}, A. 2000, \aap, 363, 1081

\bibitem[\protect\citeauthoryear{{Collier Cameron} et~al.}{{Collier Cameron}
  et~al.}{2009}]{2009MNRAS.400..451C}
{Collier Cameron}, A. et~al. 2009, \mnras, 400, 451

\bibitem[\protect\citeauthoryear{{Collier Cameron} et~al.}{{Collier Cameron}
  et~al.}{2007}]{2007MNRAS.380.1230C}
{Collier Cameron}, A. et~al. 2007, \mnras, 380, 1230

\bibitem[\protect\citeauthoryear{{Enoch} et~al.}{{Enoch}
  et~al.}{2010}]{2010arXiv1004.1991E}
{Enoch}, B., {Collier Cameron}, A., {Parley}, N.~R.,  and {Hebb}, L. 2010,
  ArXiv e-prints

\bibitem[\protect\citeauthoryear{{Girardi} et~al.}{{Girardi}
  et~al.}{2000}]{2000A&AS..141..371G}
{Girardi}, L., {Bressan}, A., {Bertelli}, G.,  and {Chiosi}, C. 2000, \aaps,
  141, 371

\bibitem[\protect\citeauthoryear{{Gonzalez} and {Laws}}{{Gonzalez} and
  {Laws}}{2000}]{2000AJ....119..390G}
{Gonzalez}, G. and {Laws}, C. 2000, \aj, 119, 390

\bibitem[\protect\citeauthoryear{{Gonzalez} et~al.}{{Gonzalez}
  et~al.}{2001}]{2001AJ....121..432G}
{Gonzalez}, G., {Laws}, C., {Tyagi}, S.,  and {Reddy}, B.~E. 2001, \aj, 121,
  432

\bibitem[\protect\citeauthoryear{{Gray}}{{Gray}}{2008}]{2008oasp.book.....G}
{Gray}, D.~F. 2008, {The Observation and Analysis of Stellar Photospheres}
  (Cambridge University Press)

\bibitem[\protect\citeauthoryear{{Hebb} et~al.}{{Hebb}
  et~al.}{2009}]{2009ApJ...693.1920H}
{Hebb}, L. et~al. 2009, \apj, 693, 1920


\bibitem[\protect\citeauthoryear{{Kurucz} and {Bell}}{{Kurucz} and
  {Bell}}{1995}]{1995all..book.....K}
{Kurucz}, R.~L. and {Bell}, B. 1995, {Kurucz CD-ROM 23: Atomic line list}
  (Smithsonian Astrophysical Observatory, Cambridge, MA)

\bibitem[\protect\citeauthoryear{{Magain}}{{Magain}}{1984}]{1984A&A...134..189%
M}
{Magain}, P. 1984, \aap, 134, 189

\bibitem[\protect\citeauthoryear{{Mandushev} et~al.}{{Mandushev}
  et~al.}{2007}]{2007ApJ...667L.195M}
{Mandushev}, G. et~al. 2007, \apjl, 667, L195

\bibitem[\protect\citeauthoryear{{Mardling}}{{Mardling}}{2007}]{2007MNRAS.382.%
1768M}
{Mardling}, R.~A. 2007, \mnras, 382, 1768

\bibitem[\protect\citeauthoryear{{McCullough} et~al.}{{McCullough}
  et~al.}{2005}]{2005PASP..117..783M}
{McCullough}, P.~R., {Stys}, J.~E., {Valenti}, J.~A., {Fleming}, S.~W.,
  {Janes}, K.~A.,  and {Heasley}, J.~N. 2005, \pasp, 117, 783

\bibitem[\protect\citeauthoryear{{O'Donovan}, {Charbonneau}, and
  {Hillenbrand}}{{O'Donovan} et~al.}{2006}]{2006AAS...20922602O}
{O'Donovan}, F.~T., {Charbonneau}, D.,  and {Hillenbrand}, L. 2006, in Bulletin
  of the American Astronomical Society, Vol.~38, Bulletin of the American
  Astronomical Society, p. 1212

\bibitem[\protect\citeauthoryear{{Pollacco} et~al.}{{Pollacco}
  et~al.}{2008}]{2008MNRAS.385.1576P}
{Pollacco}, D. et~al. 2008, \mnras, 385, 1576

\bibitem[\protect\citeauthoryear{{Pollacco} et~al.}{{Pollacco}
  et~al.}{2006}]{2006PASP..118.1407P}
{Pollacco}, D.~L. et~al. 2006, \pasp, 118, 1407

\bibitem[\protect\citeauthoryear{{Queloz} et~al.}{{Queloz}
  et~al.}{2001}]{2001A&A...379..279Q}
{Queloz}, D. et~al. 2001, \aap, 379, 279

\bibitem[\protect\citeauthoryear{{Santos}, {Israelian}, and {Mayor}}{{Santos}
  et~al.}{2004}]{2004A&A...415.1153S}
{Santos}, N.~C., {Israelian}, G.,  and {Mayor}, M. 2004, \aap, 415, 1153

\bibitem[\protect\citeauthoryear{{Sato} et~al.}{{Sato}
  et~al.}{2005}]{2005ApJ...633..465S}
{Sato}, B. et~al. 2005, \apj, 633, 465

\bibitem[\protect\citeauthoryear{{Sestito} and {Randich}}{{Sestito} and
  {Randich}}{2005}]{2005A&A...442..615S}
{Sestito}, P. and {Randich}, S. 2005, \aap, 442, 615

\bibitem[\protect\citeauthoryear{{Showman} and {Guillot}}{{Showman} and
  {Guillot}}{2002}]{2002A&A...385..166S}
{Showman}, A.~P. and {Guillot}, T. 2002, \aap, 385, 166

\bibitem[\protect\citeauthoryear{{Skrutskie} et~al.}{{Skrutskie}
  et~al.}{2006}]{2006AJ....131.1163S}
{Skrutskie}, M.~F. et~al. 2006, \aj, 131, 1163

\bibitem[\protect\citeauthoryear{{Smalley}}{{Smalley}}{2005}]{2005MSAIS...8..1%
30S}
{Smalley}, B. 2005, Memorie della Societa Astronomica Italiana Supplement, 8,
  130

\bibitem[\protect\citeauthoryear{{Smalley}, {Smith}, and {Dworetsky}}{{Smalley}
  et~al.}{2001}]{uclsyn}
{Smalley}, B., {Smith}, K.~C.,  and {Dworetsky}, M.~M. 2001, {UCLSYN User
  guide}

\bibitem[\protect\citeauthoryear{{Smith}}{{Smith}}{1992}]{1992PhDT.........1S}
{Smith}, K. 1992, Ph.D. thesis, University College, London

\bibitem[\protect\citeauthoryear{{Stetson}}{{Stetson}}{1987}]{1987PASP...99..1%
91S}
{Stetson}, P.~B. 1987, \pasp, 99, 191

\bibitem[\protect\citeauthoryear{{Torres}, {Andersen}, and
  {Gim{\'e}nez}}{{Torres} et~al.}{2010}]{2010A&ARv..18...67T}
{Torres}, G., {Andersen}, J.,  and {Gim{\'e}nez}, A. 2010, \aapr, 18, 67

\bibitem[\protect\citeauthoryear{{Wilson} et~al.}{{Wilson}
  et~al.}{2008}]{2008ApJ...675L.113W}
{Wilson}, D.~M. et~al. 2008, \apjl, 675, L113

\bibitem[\protect\citeauthoryear{{Zwitter} et~al.}{{Zwitter}
  et~al.}{2008}]{2008AJ....136..421Z}
{Zwitter}, T. et~al. 2008, \aj, 136, 421

\end{thebibliography}

\end{document}